\documentclass[aps,prl,reprint,superscriptaddress,showpacs,amsmath,amssymb]{revtex4-1}
\usepackage{graphicx}% Include figure files
\usepackage{siunitx}

\begin{document}

%\preprint{}

\title{Mode matching in second order susceptibility metamaterials}



\author{S\'ebastien H\'eron}
\affiliation{MINAO - ONERA, The French Aerospace Lab, 91761 Palaiseau, France}
\affiliation{MINAO - Laboratoire de Photonique et de Nanostructures (LPN-CNRS),
Universit\'e Paris-Saclay, Route de Nozay, 91460 Marcoussis, France}

\author{Patrick Bouchon}
\email {patrick.bouchon@onera.fr}
\affiliation{MINAO - ONERA, The French Aerospace Lab, 91761 Palaiseau, France}

\author{Riad Ha{\"i}dar}
\affiliation{MINAO - ONERA, The French Aerospace Lab, 91761 Palaiseau, France}
\affiliation{\'Ecole Polytechnique, D\'epartement de Physique, Universit\'e Paris-Saclay, 91128 Palaiseau, France}

\date{\today}

\begin{abstract}
We present an effective model for a subwavelength periodically patterned metallic layer,
its cavities being filled with a nonlinear dielectric material, which accounts for both the linear and second order behavior. 
The effective non linear susceptibility for the homogenized layer is driven by 
the nonlinearity of the dielectric material and by the geometrical parameters, thus leading
to much higher susceptibility than existing materials. This leads to 
a huge enhancement of non linear processes when used together with resonances.
Furthermore, multiple resonances are taking place in the metallic cavities,
and we investigate the mode matching situations for frequency conversion processes 
and show how it enhances further their efficiency.
\end{abstract}

%\keywords{}

\maketitle

Metamaterials are artificial materials, obtained with 
subwavelength patterned elements, that exhibit
effective electromagnetic properties which depend not only 
on the material, but also on the geometry.
They have given birth to original and unprecedented behaviors
both in the linear and non linear regime, such as 
optical cloaking, phase matched negative index or
left-handed metamaterials
\cite{zharov2003nonlinear,cai2007optical,rose2011controlling}.
Subwavelength patterned elements can behave as nanoantennas
able to funnel the incoming light and concentrate it in 
a small volume, which is extremely appealing in 
the context of non linear optics \cite{kauranen2012nonlinear,czaplicki2013enhancement,lapine2014colloquium,lee2014giant,minovich2015functional}. Indeed, optical nanoantennas can 
provide huge enhancement of the electric field, and even if 
the volume at stake is smaller compared to the whole device, 
nonlinear effects can be boosted. 
Most of the nanoantennas reported in the literature are metallic, 
as they can confine the field more easily than dielectric antenna.
So the surface nonlinearities of the metal itself are enhanced \cite{czaplicki2013enhancement,wang2009surface,genevet2010large,thyagarajan2013augmenting},
even if dielectric materials can provide much higher volume nonlinearities.

Besides, there are two important stakes when considering a non linear metamaterial. 
The first is to predict the linear and non linear properties 
of the patterned material by its geometrical parameters. 
It can be done for instance with the Maxwell Garnett formalism \cite{PhysRevA.46.1614}, 
through field averaging \cite{smith2006homogenization}
or by retrieving it from rigorous computations or experiments \cite{larouche2010retrieval}.
The second stake consists in finding metamaterials exhibiting multiple resonances 
so as to enhance the field at each one of the wavelength involved in the 
frequency conversion process. 
Lately, several metallic mode matching nanostructures have been 
suggested to further improve the efficiency of non linear effects 
either based on plasmonic nanoantennas \cite{Park:12,celebrano2015mode} or on
phased-array sources \cite{wolf2015phased}.

\begin{figure}[ht]
\centering
  \includegraphics[width=0.48\textwidth]{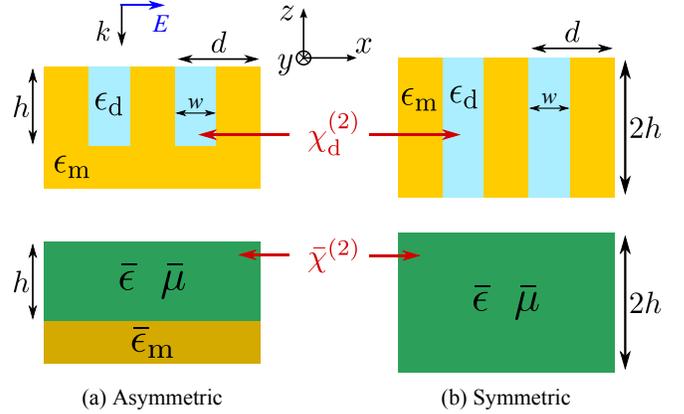}
  \caption{(a) Periodic grating (period $d$) of metallic grooves 
  of width $w$ and height $h$, filled with a non linear dielectric 
  (permittivity $\epsilon_\mathrm{d}$, non linear susceptibility $\chi^{(2)}_\mathrm{d}$.)
  The waves are normally incident and TM polarized with wave vectors $\mathbf{k}$ lying in the $xOz$ plane.
  Below is shown the equivalent metamaterial, that consists in an homogeneous layer with 
  effective permittivity, permeability and non linear susceptibility.
  (b) Periodic grating of metallic slits filled with a non linear dielectric, 
  which parameters are similar to the reflecting case.
  }
  \label{fig:fig1}
\end{figure}

In this letter, we report on mode matching in a high susceptibility 
metamaterial for frequency conversion. The investigated structure consists 
in a patterned metallic layer, filled with a non linear dielectric, 
that melts the high confinement properties of metallic nanoantennas
and the high non linear susceptibility of chosen dielectric materials. 
It additionally exhibits multiple Fabry-Perot resonances that can be used for mode matching. 
First, we present 
an effective model that fairly accounts for both linear and non linear 
behaviors of the structure. Its effective linear and non linear optical properties 
are mainly determined by the aperture ratio. One of the main differences 
with previously studied plasmonic structures
lies in the monitoring of the non linear response by the material 
filling the holes in the metallic layer rather than the
metallic surface generation itself. 
Then, we show how mode matching can be achieved in the case of second harmonic 
generation (SHG) and difference frequency generation (DFG), allowing to 
reach higher conversion efficiency. These results are scalable to large 
spectral ranges, and can be adapted in the context of metasurfaces based on MIM antennas.

We aim at describing a subwavelength periodic metal-dielectric layer as an effective medium 
where the dielectric inclusions display a second order non linear susceptibility. 
Two configurations of this layer are considered, as shown in Fig. \ref{fig:fig1}. 
In the one case, the metal-dielectric layer is placed upon a metallic substrate (grating of grooves) and acts as a reflection device
and in the other case, the metal-dielectric layer is surrounded by air (grating of slits).
For the sake of simplicity, the permittivity of the metal $\epsilon_\mathrm{m}$ is considered identical in the layer and the substrate,
while the dielectric inclusions bear a permittivity $\epsilon_\mathrm{d}$ and a non linear susceptibility tensor $\chi^{(2)}_\mathrm{d}$ which contains only $\chi^{(2)}_{iii}$ terms.
The incoming wave is normally incident and transverse magnetic (TM) polarized, 
at the wavelength $\lambda$ with a wave vector $k_0 = 2\pi/\lambda$. 
 The period of the system is $d$ and is subwavelength, the grooves or slits have a height $h$ and a width $w$. 
The transmission case (Fig. \ref{fig:fig1} (b)) has previously been described as a metamaterial for perfect metals
and was involving an effective thickness \cite{shen2005mechanism,ironside2013orders}.
One of the first challenge is to take a lossy metal into account, and to describe the effective layer 
through the sole effective optical properties $\bar \epsilon$, $\bar \mu$ and $\bar \chi ^{(2)}$.

First, the normalized wave vector of the fundamental mode $\sqrt{\epsilon_\mathrm{TM}}$ propagating
in the plane waveguide set by the two metallic surfaces obeys to the equation:
\begin{equation}
\tanh \left( \sqrt{ \epsilon_\mathrm{TM} - \epsilon_\mathrm{d} } \frac{w}{2} \right) = - \frac{\epsilon_\mathrm{d}}{\epsilon_\mathrm{m}} 
\sqrt{ \frac{ \epsilon_\mathrm{TM} - \epsilon_\mathrm{m} }{  \epsilon_\mathrm{TM} - \epsilon_\mathrm{d} } }.
\end{equation}
After some tedious calculations, this equation can be solved at 
the first order since $\epsilon_\mathrm{d} \ll \epsilon_\mathrm{m}$, and is written as:
\begin{equation}
\epsilon_\mathrm{TM} = \epsilon_\mathrm{d} \left( 1 + \frac{2 \delta}{w} - \frac{\epsilon_\mathrm{d} }{\epsilon_\mathrm{m}}  \right),
\label{eq:epsTM}
\end{equation}
 where $\delta = i \lambda / 2 \pi \sqrt{\epsilon_\mathrm{m}}$ is the metal skin depth. 

 The light incoming onto the structure is either reflected or funneled into the slit \cite{pardo2011light,bouchon2011total}, 
 so that the energy in the metal-dielectric layer is contained in the dielectric inclusions. 
Consequently, the stored energy is the same in both the effective layer and in the dielectric inclusions:
\begin{equation}
\int_{z=0}^{z= h} \int_{x=0}^{x=d} \bar{\mathbf{E}}.\bar{\mathbf{D}} = \int_{z=0}^{z=h} \int_{x=0}^{x=w} \mathbf{E}.\mathbf{D},
\label{eq:ener}
\end{equation}
where $\mathbf{D}$ is the electric displacement field, and $\bar{\mathbf{E}}$ and $\bar{\mathbf{D}}$ stand for the fields value in the effective layer . It must be 
emphasized that the bounds of integration along $x$ have been limited 
to the dielectric since the energy stored in the metallic sidewalls is 
negligible. Indeed, at the metal-dielectric interface normal to the $x$ axis, 
the $x$ component of the electric field is discontinuous, and the normal 
electric field on each side are linked by:
\begin{equation}
\frac{E_x(x=w^-)}{E_x(x=w^+)} = \frac{\epsilon_\mathrm{m}}{\epsilon_\mathrm{d}} \gg 1.
\label{eq:cont}
\end{equation}
Thus, the amplitude of the electric fields inside the dielectric inclusion is far greater than inside the metal. 

We consider that the fundamental guided mode is phase and amplitude 
invariant along the $x$ direction, so that Eq. \ref{eq:ener} is expressed as:
\begin{equation}
d \times \int_{z=0}^{z= h} \bar \epsilon \bar E^2 = w \times \int_{z=0}^{z=h} \epsilon_\mathrm{d} E^2.
\end{equation}
Besides, this equation is valid for all $h$, so it can be further simplified to $\bar E^2 d \bar \epsilon  = E^2 w \epsilon_\mathrm{d}$.
The potential difference inside one period, has to be equal between 
the original configuration and the effective one, so that $\bar E d= E w $. 
The effective permittivity is then obtained as:
\begin{equation}
\bar \epsilon = \epsilon_\mathrm{d} \times \frac{d}{w}.
\label{eq:effective_eps}
\end{equation}

Eventually, the phase accumulated by a wave during its propagation through the structure is the same in both cases, $k h = \bar k h$. 
It writes as $ \bar \epsilon \bar \mu = \epsilon_\mathrm{TM}$ where the effective layer is chosen magnetic, and its effective permeability 
can be expressed thanks to Eq. \ref{eq:effective_eps}:
\begin{equation}
\bar \mu = \frac{\epsilon_\mathrm{TM}}{\epsilon_\mathrm{d}} \times \frac{w}{d}.
\end{equation}

In the asymmetric case (see Fig. \ref{fig:fig1}(b)), the equivalent 
layer has to be deposited on a mirror which displays an effective 
permittivity $\bar \epsilon_\mathrm{m}$ different from the one of the metal. 
It can be expressed by matching the reflection coefficients at the bottom of the slit:
\begin{equation}
\frac{ \sqrt{ \bar \epsilon / \bar \mu} - \sqrt{ \bar \epsilon_\mathrm{m}} }{ \sqrt{\bar \epsilon / \bar \mu} + \sqrt{ \bar \epsilon_\mathrm{m}} } = \frac{ \sqrt{ \epsilon_\mathrm{TM}} - \sqrt{ \epsilon_\mathrm{m}} }{ \sqrt{\epsilon_\mathrm{TM}} + \sqrt{ \epsilon_\mathrm{m}} }.
\end{equation}
The effective permittivity of the metallic substrate is given by $\bar \epsilon_\mathrm{m} = \epsilon_\mathrm{m} \// \bar \mu^2$.

In the following, the asymmetric resonator has a 
period $d = \si{1 \, \micro \meter}$, a width $w = \si{ 0,1 \, \micro \meter}$ 
and a height $h = \si{0,5 \, \micro \meter}$. The metal is gold, described 
by a Drude model fitting Palik data \cite{olmon2012optical} and the dielectric is gallium arsenide 
which optical properties are taken from the literature \cite{skauli2003improved}.
All the parameters are identical for the symmetric case apart for the height $h = \si{ 1 \, \micro \meter}$. 
The computations are performed with the B-spline modal method, which makes a fast and exact resolution of Maxwell equations, 
and can also solve the nonlinear behavior under the undepleted pump approximation \cite{bouchon2010bmm,heron2015modal}.

The linear response of both structures shows(see Supplemental Materials \cite{EPAPS})
a rather fair agreement obtained with the effective metamaterial. 
As expected, these layers induce Fabry-Perot resonances leading to reflectivity dips and transmittivity peaks, 
at wavelengths determined by solving the phase condition inside the effective layer:
\begin{equation}
\lambda _m = \frac{2 \sqrt{\epsilon_\mathrm{TM}} h^* }{ m - \phi/2\pi}, 
\label{eq:FP}
\end{equation}
where $m \in \mathbb{N^*}$, and 
$\phi$ is the phase of the bottom reflection coefficient. It is equal to 
zero for the symmetric situation and to $\pi$ for the asymmetric one. 
To take into account the penetration of the propagating mode in the bottom metal in the asymmetric case, an equivalent height $h^* = h+ \delta$ is introduced 
in the asymmetric case and $h^*=h$ in the symmetric one. 
Streamlines of the Poynting vector at the resonance wavelengths are also
shown in Supp. Mat. \cite{EPAPS} to illustrate the funneling phenomenon, that was used 
in Eq. \ref{eq:ener}.

\begin{figure}[ht]
\centering
  \includegraphics[width=0.48\textwidth]{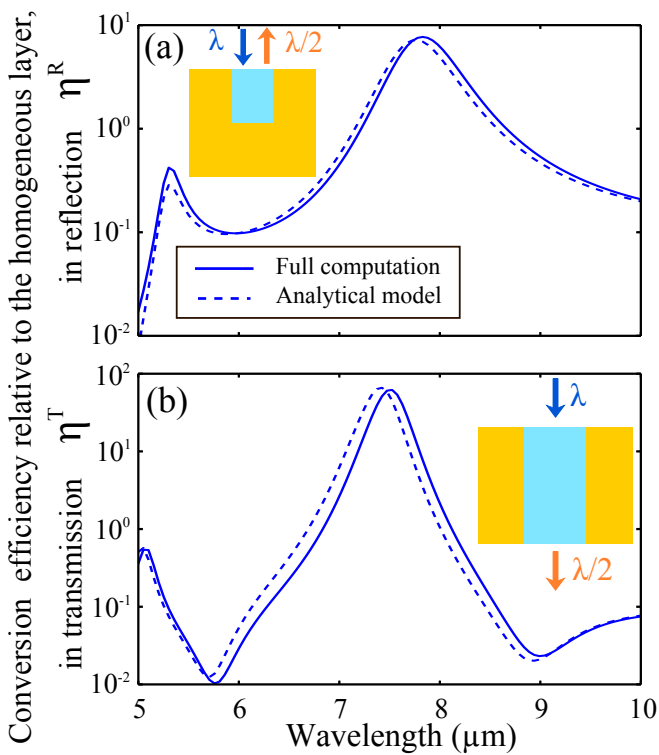}
  \caption{SHG intensity ratio between structured 
  and unstructured layers: (a) in reflection for the asymmetric case and 
  (b) in transmission for the symmetric one, as a function of the wavelength. 
  Continuous lines stand for the full computation, whereas dashed ones stand 
  for the analytical model. Involved parameters are: $d = \si{1 \, \micro\meter}$, 
  $w = \si{0,2 \, \micro\meter}$, $h = \si{0,5 \, \micro\meter}$ for asymmetric case 
  or $h = \si{1 \, \micro\meter}$ for symmetric case, $\chi^{(2)} = \si{150 \, \pico \meter \per\volt}$.}
  \label{fig:fig2}
\end{figure}

Up to this point, the linear characteristics of the effective layer have 
been fully determined, but this layer also behaves as a medium with a 
higher non linear susceptibility. In order to determine its effective value, 
we state that the nonlinear part of the electromagnetic energy stored in 
one period is the same in the two cases, as it was written in Eq. \ref{eq:ener} for 
the linear part of the stored energy. This term is proportional to $\mathbf{E} . \mathbf{P^{(2)}}$, 
and using the same arguments for integration than before, it leads to:
\begin{equation}
\bar \chi^{(2)} \bar E_{\lambda_{1}} \bar E_{\lambda_{2}} \bar E_{\lambda_{3}} \times d = \chi^{(2)}_\mathrm{d} E_{\lambda_{1}} E_{\lambda_{2}} E_{\lambda_{3}} \times w,
\label{eq:NLenergie}
\end{equation}
where $\lambda _1$ and $\lambda _2$ are the pumps wavelengths, and $\lambda _3$ the signal wavelength. 

The effective nonlinear susceptibility is eventually found to be:
\begin{equation}
\frac{\bar \chi^{(2)}}{\chi^{(2)}_\mathrm{d}} =   \left( \frac{d}{w} \right)^2 
\label{eq:chi2}
\end{equation}
It illustrates the great enhancement of the quantity of nonlinear sources 
inside the cavity of such structures, as $d/w$ is higher than one. For instance, 
in the two examples of Fig. \ref{fig:fig2}, the effective non linear susceptibility is 
increased by two orders of magnitude. 
However, the non linear susceptibility is not the only parameter involved in the 
efficiency of frequency conversion processes. In fact, due to the high value of the effective
permittivity, for most of the wavelengths there is no impedance matching. 
So the incoming wave is not penetrating the non linear 
metamaterial, which results in a poor efficiency of the non linear processes.

The efficiency of the second harmonic generation is computed for both structures in Fig. \ref{fig:fig2}.
For the sake of comparison, the plotted efficiency is normalized by the intensity of a non 
patterned gallium arsenide layer, which thickness is chosen so as 
to display Fabry Perot resonances at the same wavelengths. Following Eqs. (\ref{eq:epsTM},\ref{eq:FP}), the 
equivalent GaAs layer is a bit thicker than the patterned layer.
The relative conversion efficiency is defined in reflection
as $\eta ^{R} = \frac{I^R_\mathrm{out} }{I^R_\mathrm{out,ref}}$
where $I^R_\mathrm{out}$ is the output reflected
nonlinear intensity of the metamaterial, 
and $I^R_\mathrm{out,ref}$ is the reflected non linear intensity 
for an homogeneous layer of gallium arsenide exhibiting a fundamental Fabry Perot 
resonance at the same wavelength than the metamaterial layer (see Eq. \ref{eq:FP}).
Due to the Eq. \ref{eq:epsTM}, the gallium arsenide layer is slightly thicker 
than the metamaterial layer.
The relative conversion
efficiency in transmission $ \eta ^T$ is defined in a similar way. 
The GaAs nonlinear susceptibility is chosen as $\chi^{(2)} = \si{150 \, \volt \per \pico \meter}$. 
The full computation for the patterned layer is plotted in continuous lines, 
while dashed ones stand for the metamaterial model. 

\begin{figure*}[htb]
\centering
  \includegraphics[width=0.88\textwidth]{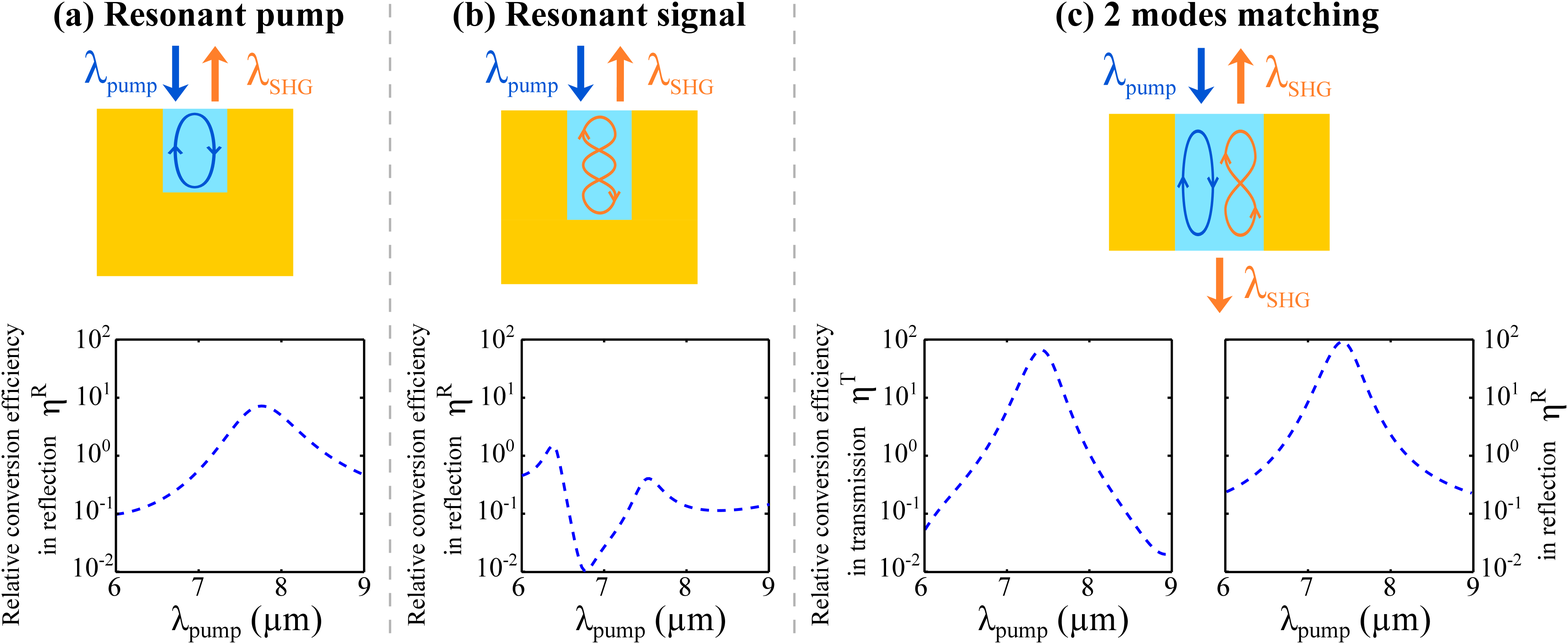}
  \caption{Three scenarii of resonant behaviours for SHG in the metal-dielectric layer: 
  (a) resonant pump ($h = \si{0,5 \, \micro\meter}$, $h _{\mathrm{GaAs}} = \si{0,55 \, \micro\meter} $) or 
  (b) resonant signal  ($h = \si{0,73 \, \micro\meter}$, $h _{\mathrm{GaAs}} = \si{0,82 \, \micro\meter} $) 
  in an asymmetric resonator, 
  and (c) both resonant pump and signal  ($h = \si{1 \, \micro\meter}$, 
  $h _{\mathrm{GaAs}} = \si{1,15 \, \micro\meter} $) in a symmetric resonator 
  creating a mode matching configuration. The efficiency curves are 
  shown below as functions of the pump wavelength. 
  In the three cases, $d = \si{1 \, \micro\meter}$ and $w = \si{0,2 \, \micro\meter}$. 
  }
  \label{fig:fig3}
\end{figure*}

Two noticeable behaviors corresponding to resonant and non-resonant cases appear. 
First, the maximum of the second harmonic signal is indeed one order of magnitude greater leading 
to interesting resonant values of the enhancement. 
Second, the ratio drops below 1 meaning that non-resonant behavior gives worse 
results for the structured resonators. This stems from the small value of the 
transmission coefficients at the interface in the structured case as the impedance 
$Z = \sqrt{ \bar \mu / \bar \epsilon}$ reaches huge values in this case. 
The second harmonic light is hardly driven to the outer medium compared to the homogeneous layer case, 
leading to poor values of efficiency away from the resonance. 
Interestingly, the symmetric case leads to a better conversion efficiency both in 
reflection (data not shown) and transmission.
In fact, this is a direct consequence of the presence of harmonics resonance at wavelengths given by Eq. (\ref{eq:FP}),
which may result in mode matching situations where both the pump wavelength and the SHG signal are subject to a resonance.
The various scenarii of resonant behaviors in both structures for SHG or DFG are investigated below.

Figure \ref{fig:fig3} shows the 3 resonant situations that happen in the case of SHG
with the respective conversion efficiency spectra. 
On the one hand, the incoming pump wave at wavelength $\lambda _{\mathrm{pump}}$ can be resonant 
to increase the quantity of created nonlinear polarization (Fig. \ref{fig:fig3} (a)). 
On the other hand, the outcoming signal wave at $\lambda _{\mathrm{SHG}}$ can be resonant to enhance 
the coupling from the slit to the outer medium (Fig. \ref{fig:fig3} (b)). 
When both conditions are fulfilled, it is a mode matching situation 
(Fig. \ref{fig:fig3} (c)) where the nonlinear intensity ratio reaches its highest 
value for a selected period. In the asymmetric resonator, only the cases of 
Fig. \ref{fig:fig3} (a) and (b) can happen, thus limiting the value of $\eta$ to the 
one obtained when the pump is solely resonant. Using a resonance at the second harmonic 
wavelength is typically one order of magnitude less efficient, since the 
energy generated at the second harmonic depends only linearly on the second harmonic electric field (see Eq. \ref{eq:NLenergie}).
In the case of the symmetric resonator, there is always a mode matching situation for SHG between the fundamental
resonance at $\lambda$ and the first order of resonance at $\lambda /2$. The low quality factors of both resonances 
can compensate for the natural dispersivity of the material.

\begin{figure}[ht]
\centering
  \includegraphics[width=0.45\textwidth]{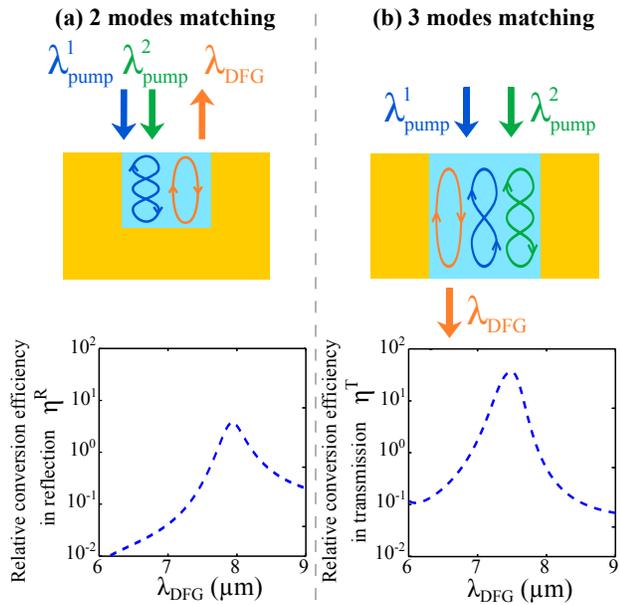}
  \caption{Two scenarii of modes matching for DFG: 
  (a) two modes matching with one resonant pump and a resonance at the 
  DFG signal in an asymmetric resonator ($h = \si{0,5 \, \micro\meter}$, $h _{\mathrm{GaAs}} = \si{0,55 \, \micro\meter} $), and 
  (b) three modes matching (both resonant pumps and resonant DFG signal)
   in a symmetric resonator ($h = \si{1 \, \micro\meter}$, $h _{\mathrm{GaAs}} = \si{1,15 \, \micro\meter} $). 
   The efficiency curves are shown below as a function of the pump wavelength. 
   The other parameters are the same in both structures ($d = \si{1 \, \micro\meter}$, $w = \si{0,2 \, \micro\meter}$).
   }
  \label{fig:fig4}
\end{figure}

There are various configurations of mode matching in both structures for DFG. 
Two of them are illustrated in the case of the asymmetric (resp. symmetric) resonator 
in Fig. \ref{fig:fig4}(a) (resp. Fig. \ref{fig:fig4}(b)).
In both resonators, there is a degree of freedom to choose the wavelengths 
in order to be in a two modes matching situation. For instance, the signal wavelength $\lambda _{\mathrm{DFG}}$
determines the geometry of the resonator, and one pump wavelength is chosen 
so as to match one of the harmonics of the resonator
while the last one is determined by the energy conservation condition. 
The conversion efficiency shown in Fig. \ref{fig:fig4}(a) is comparable 
to the one obtained for SHG in Fig. \ref{fig:fig3}(a), which is explained by 
the fact that the pump is degenerate so it could be considered as a degenerate 2 mode matching
configuration.
Following this previous scheme, three modes matching can be 
straightforwardly obtained in the symmetric resonator.
Indeed, Eq. \ref{eq:FP} quantifies the energy of each harmonic wavelength as a multiple of 
the fundamental wavelength energy. So apart from some peculiar cases, 
if two of the wavelengths involved in the DFG process have
been chosen at resonance wavelengths, the third one is also at another 
resonance wavelength due to the energy conservation condition (and neglecting the dispersivity). 
In Fig. \ref{fig:fig4}(b), the fundamental wavelength as well as the two first harmonics wavelengths 
are used ($\lambda _{\mathrm{pump}} ^1 =\lambda _{\mathrm{DFG}} /3$ and $\lambda _{\mathrm{pump}} ^2 =\lambda _{\mathrm{DFG}} /2$).  
As expected, it leads to a higher efficiency conversion ratio than in the two mode matching situation
for both transmission and reflection (data not shown). However, 
this enhancement is lower than for the SHG , due to the fact that the
natural dispersivity of the gallium arsenide must be managed for three different 
wavelengths. 
To conclude, non linear phenomena in subwavelength 
metallic slits or grooves filled with a nonlinear material can be fairly 
described by this homogenization model.
This metamaterial exhibits an unusually high nonlinear effective susceptibility that leads to 
higher efficiency of the frequency conversion processes, which can be even further enhanced by exploiting
mode matching between resonances. 
It must be emphasized that the metamaterial properties can be spatially tuned, by simply changing the in-plane 
geometrical parameters, making it possible for instance to address various wavelength ranges. 
These results can be directly applied to various metals and non linear dielectric materials. In the mid infrared range,
the efficiency for thick layer of metamaterials is plagued by the ohmic metallic losses, 
but it is no longer the case for higher wavelength ranges.

\begin{acknowledgments}
We acknowledge financial support from the ONERA through the MOLIERE project and from a DGA-MRIS scholarship.
\end{acknowledgments}

%
%\bibitem{EPAPS}%
%  \BibitemOpen
%  \bibinfo {note} {See Supplemental material at [URL] for Fig. 1 which validate the analytical model of the Metamaterial.}%

\end{document}